\documentclass[aps,pra,reprint,groupedaddress,showpacs,showkeys]{revtex4-1}
\usepackage{amsmath,graphicx,hyperref}
\usepackage[FIGTOPCAP,raggedright,nooneline,bf,footnotesize]{subfigure}
\usepackage{indentfirst}
\usepackage{braket}
\usepackage{diagbox}
\usepackage{amssymb}
\usepackage{bbold}
\usepackage{colortbl}
\usepackage{multirow}

\newcommand{\figref}[1]{Fig.~\ref{#1}}
\newcommand{\tabref}[1]{Tab.~\ref{#1}}
\renewcommand{\eqref}[1]{Eq.~\ref{#1}}

\newcommand{\tr}{{\mathrm{Tr}}}
\newcommand{\dd}{{\mathrm{d}}}
\newcommand{\prop}{\mathcal{S}}
\newcommand{\centered}[1]{\begin{tabular}{l} #1 \end{tabular}}

\begin{document}
	
	\title{Tomography of time-bin quantum states using time-resolved detection}
	
	\author{Karolina Sedziak-Kacprowicz}
	\email[]{These authors contributed equally to this work.}
	\author{Artur Czerwinski}
	\email[]{These authors contributed equally to this work.}
	\author{Piotr Kolenderski}
	\email{kolenderski@fizyka.umk.pl}
	\affiliation{Institute of Physics, Faculty of Physics, Astronomy and Informatics, Nicolaus Copernicus University, Grudziadzka 5, 87-100 Torun, Poland} 
%\keywords{quantum communication; quantum state tomography; time-domain encoding; telecommunication fibers}
	
	\begin{abstract}
	 	We present a method for measuring quantum states encoded in the temporal modes of photons. The basis for the multilevel quantum states is defined by the use of modes propagating in a dispersive medium, which is a fiber in this case. The propagation and time-resolved single photon detection allow us to define a positive-operator valued measure (POVM). The POVM depends on the amount of dispersion and the characteristics of a detector. This framework is numerically tested by performing quantum state tomography on a large number of states for a set of realistic experimental settings. Finally, the average fidelity between the expected and reconstructed states is computed for qubits, qutrits and entangled qubits.
	\end{abstract}
	\maketitle
	
	\section{Introduction}
	
	Quantum communication protocols can be implemented using photonic states. A generic scheme is based on transmission of a single photon through a channel to a receiver. Information can be encoded, for example in polarization \cite{Clausen2014}, spatial \cite{Lima2009,Solis-Prosser2013} or spectral modes \cite{Reimer2016}. The channel properties and the receiver characteristics determine the optimal protocol assuring the best performance. Free space \cite{Yin2017,Pugh2017,Liao2018} and fiber-based \cite{Gisin2002} quantum communication links have been demonstrated. Each of them has its advantages and disadvantages. Both suffer from several effects limiting their maximal distance and throughput. A typical fiber introduces uncontrollable polarization transformation, which must be taken into account \cite{Banaszek2004}, when a qubit is encoded in polarization. Fiber links are also subjected to dispersion and loss, which, when combined with imperfect detectors, limits the transmission maximal range \cite{Sedziak2017}. On the other hand, one can take advantage of the propagation effects in the fiber to extend the distance of quantum communication protocols \cite{Lasota2018}, when time-resolved single photon detection is available. 
	
	It is also possible to encode information in the temporal domain by using interferometric techniques. The first proposal involving time-bin encoding was introduced by Franson in the context of the violation of Bell inequalities \cite{Franson1989}. Then, this idea was successfully applied to quantum key distribution protocols \cite{Stucki2005,Inagaki2013,Ikuta2018}, quantum information processing \cite{Donohue2013, Humphreys2013} and more recently in quantum teleportation\cite{Anderson2020}. Experimental realization of time-bins requires one unbalanced interferometer to prepare time-bin states and another interferometer or a nonlinear interaction for measurement. The advantages are the noise robustness during propagation of photons and a simplified experimental setup to realize quantum communication protocol.
	
	We adapt a framework where the Hilbert space of a multilevel system, a qudit, can be established based on a discrete number of separated temporal modes of a single photon, time bins \cite{Franson1989}. The unitary evolution that the photon experiences during propagation in a dispersive medium can be interpreted as an evolution of a qudit state within the Schr{\"o}dinger picture. On the other hand, the Heisenberg representation allows us to define measurement operators which change in time. This is analogous to spatial encoding in the transverse momentum of a photon \cite{Neves2007}, where the photon propagates through a system of multiple slits that defines its state. The photon is then measured using spatially-resolved single photon detection technique, which defines a POVM. In this paper, we first introduce the framework for qubits and generalize it for qudits. Next, the method's robustness is tested numerically by analysing the quantum state tomography results for qubits, qutrits and entangled qubits. 
	
\section{Temporal encoding}
\subsection{Qubit}

	Let us assume a physical situation where a single photon state is described by a wave function that is the sum of a pair of separated temporal modes
	\begin{equation}
	\psi(t)=\alpha_0 u\left(t+\tau/2\right)+\alpha_1 u\left(t-\tau/2\right),
	\label{eq:def:qubit:cont}
	\end{equation}
	where $\alpha_0$ and $\alpha_1$ are complex numbers satisfying the normalization condition, $|\alpha_0|^2+|\alpha_1|^2=1$, and
	\begin{equation}
	u(t)=\frac{e^{ -\frac{t^2}{2 \sigma ^2}}}{\sqrt{\sqrt{\pi }} \sqrt{\sigma }}.
	\label{eq:ut}
	\end{equation}
	
	\begin{figure}[h]
		\centering
		\includegraphics[width=0.85\columnwidth]{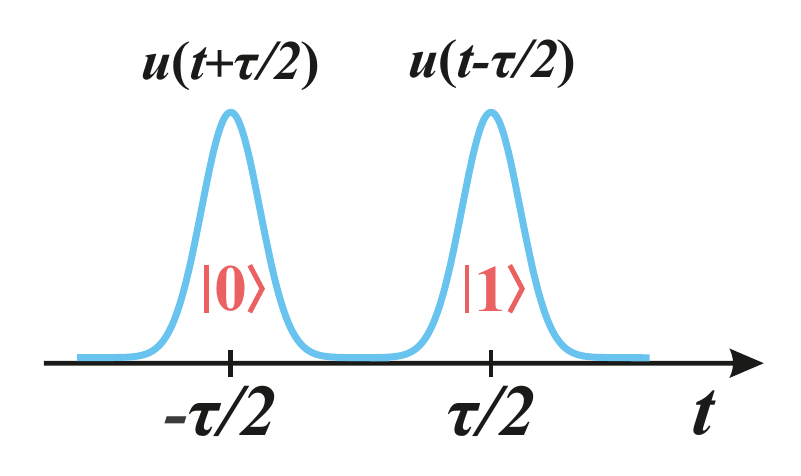}
		\caption{A single qubit is defined as a photon which is delocalized in two wave packets separated in time interval $[-\tau/2,\tau/2]$.}
		\label{fig:figf}
	\end{figure}	
	This is depicted in \figref{fig:figf}. Let us now define the following vectors
	\begin{eqnarray}
	\ket{0}&=&\int_{-\infty}^{\infty}\dd t\, u(t+\tau/2)\ket{t},\\
	\ket{1}&=&\int_{-\infty}^{\infty}\dd t\, u(t-\tau/2)\ket{t},
	\end{eqnarray}
	where $\ket{t}$ represents the state of a photon localized within a time instant $t$. The overlap of the two state vectors reads
	\begin{equation}
	\braket{0|1}=e^{\frac{-\tau^2}{4\sigma^2}},
	\end{equation}
	which means that they are not perfectly orthogonal, but can be made approximately so by a proper choice of the ratio of the modes' separation, $\tau$, to their widths $\sigma$. A realistic assumption is $\braket{0|1}=3\times 10^{-7}$  for $\tau/\sigma=7.7$. Therefore, we can consider $\{\ket{0}, \ket{1}\}$ as an orthogonal basis. This allows us to define a logic qubit state as
	\begin{equation}
	\ket{\psi_{in}}=\alpha_0\ket{0}+\alpha_1\ket{1}.
	\label{eq:qubit:def}
	\end{equation}
	This is the definition that will be generalized in the next section.
	To define the measurements on a logic qubit, let us first consider a wave packet, $u_L(t)$, propagated through a fiber characterized by its length $L$ and a dispersion parameter $\beta$. This can be modelled as an action of a propagator, $\prop (t,t',L)$ \cite{Sedziak2017}, on the initial state, $u(t')$, in the following way 
	\begin{equation}
	u_L(t)=\int_{-\infty}^{\infty}\mathrm{d}t'\,\prop (t,t',L) u(t').
	\label{eq:uL}
	\end{equation}
	For the propagator related to a dispersive fiber (see Ref.~\cite{Sedziak2017} for details) this results in
	\begin{equation}
	u_L (t)=\frac{e^{\frac{i t^2}{4 \beta  L-2 i \sigma ^2}}}{\sqrt[4]{\pi }\sqrt{ \sigma+\frac{2 i \beta  L}{\sigma } }}.
	\end{equation}
	With this result, we can apply the Born rule to compute the probability density of detecting a photon at time, $t$ after the propagation
	\begin{equation}
	p(t)= |\alpha_0 u_L\left(t+\tau/2\right)+\alpha_1 u_L\left(t-\tau/2\right)|^2.
	\label{eq:prob:dens}
	\end{equation}
	We observe that the last equation can be rewritten as
	\begin{equation}
	p(t)=\tr\left(\hat M(t)\ket{\psi_{in}}\bra{\psi_{in}}\right),
	\label{eq:br}
	\end{equation}
	assuming the measurement operator is given by
	\begin{equation}
	\hat{M}(t)=\mu(t)\ket{\psi_M(t)}\bra{\psi_M(t)}.
	\label{eq:Mt}
	\end{equation}
	with
	\begin{equation}
	\mu(t)=\frac{\sigma }{\sqrt{\pi } \sqrt{4 \beta ^2 L^2+\sigma ^4}} \left(e^{-\frac{\sigma ^2 (t + \tau/2)^2}{4 \beta ^2 L^2+\sigma ^4}}+e^{-\frac{\sigma ^2 (t-\tau/2 )^2}{4 \beta ^2 L^2+\sigma ^4}}\right),
	\end{equation}
	which is interpreted as the weight of the normalized state defined as
	\begin{equation}
	\ket{\psi_M(t)}=\frac{1}{\sqrt{\mu(t)}}\left(
	\begin{array}{c}
	u_L(t+\tau/2) \\
	u_L(t-\tau/2)\\
	\end{array}
	\right).
	\end{equation}
	Note that the operator, $M(t)$, depends only on the fiber parameters $L$ and $\beta$ and does not depend on the initial state $\ket{\psi_{in}}$. It can also be easily shown that it obeys the following relation:
	\begin{equation}
	\int_{-\infty}^{\infty} \hat{M}(t)\dd t=\begin{pmatrix} 1&e^{-\tau^2/4\sigma^2}\\e^{-\tau^2/4\sigma^2}&1\end{pmatrix} \approx \mathbb{1},
	\end{equation}
	which makes it a proper POVM with the approximation that the off-diagonal terms are negligible. The same assumptions make the basis states orthogonal.
	
\begin{figure*}[ht]
		\centering
		\begin{tabular}{c|c|c|c}
			\backslashbox[12mm]{\color{black}$L$}{\color{black}$\sigma_D$}&\centered{$0$ ps}&\centered{$1$ ps}&\centered{$4$ ps}\\\hline
			\centered{$200$ m}&\centered{{\includegraphics[width=0.45\columnwidth]{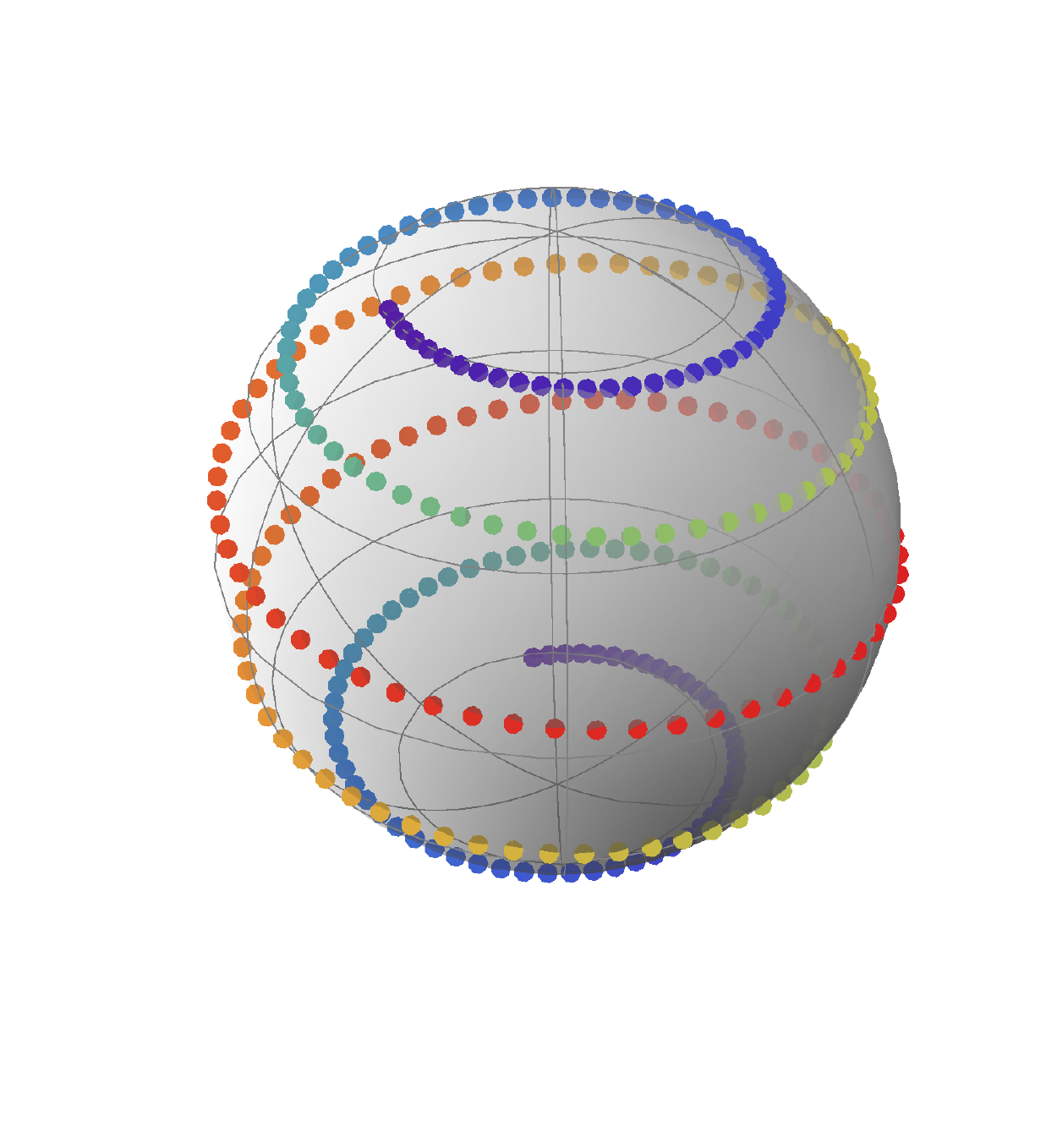}} \\
				{\includegraphics[width=0.5\columnwidth]{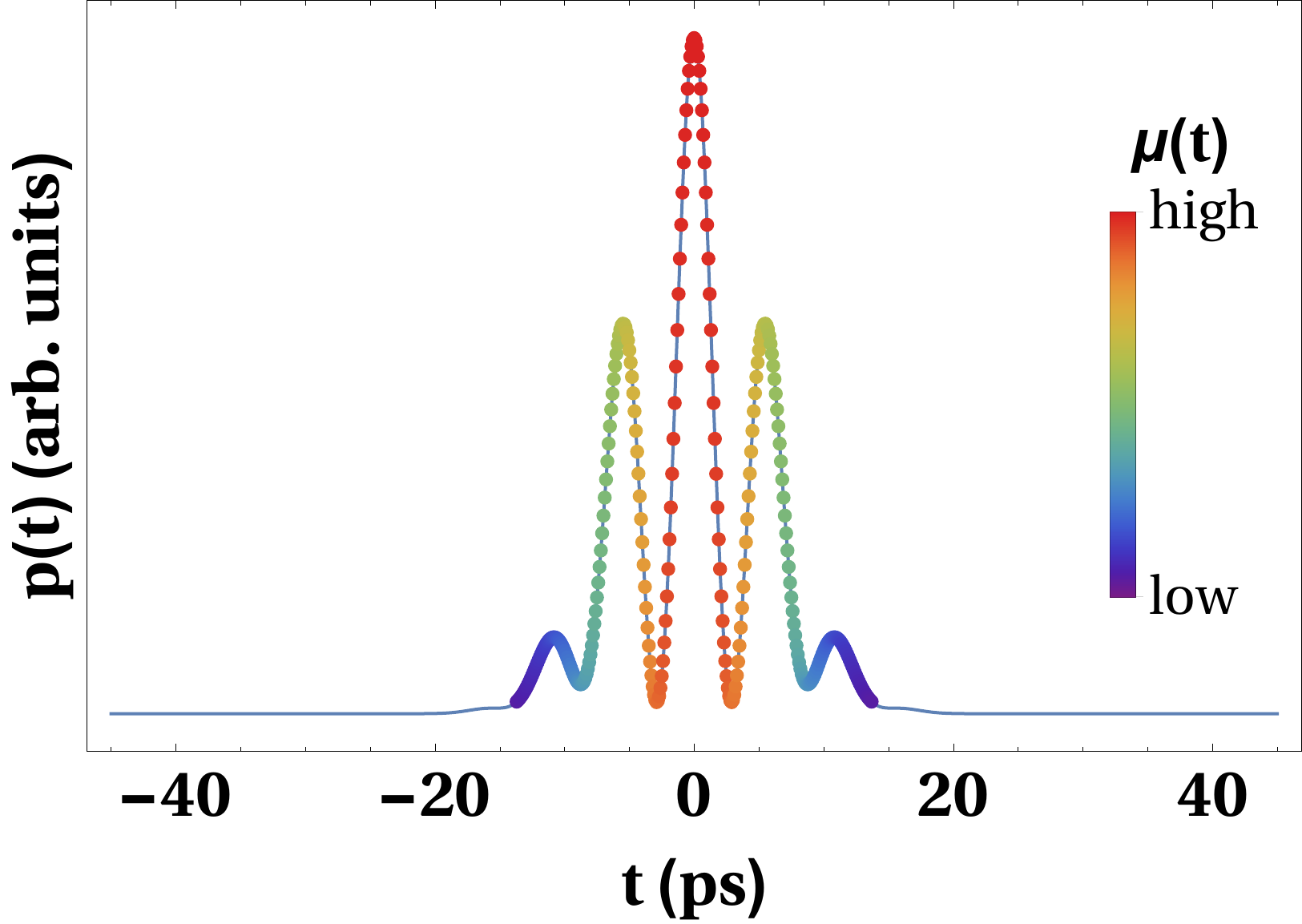}}	
			}&
			\centered{\includegraphics[width=0.45\columnwidth]{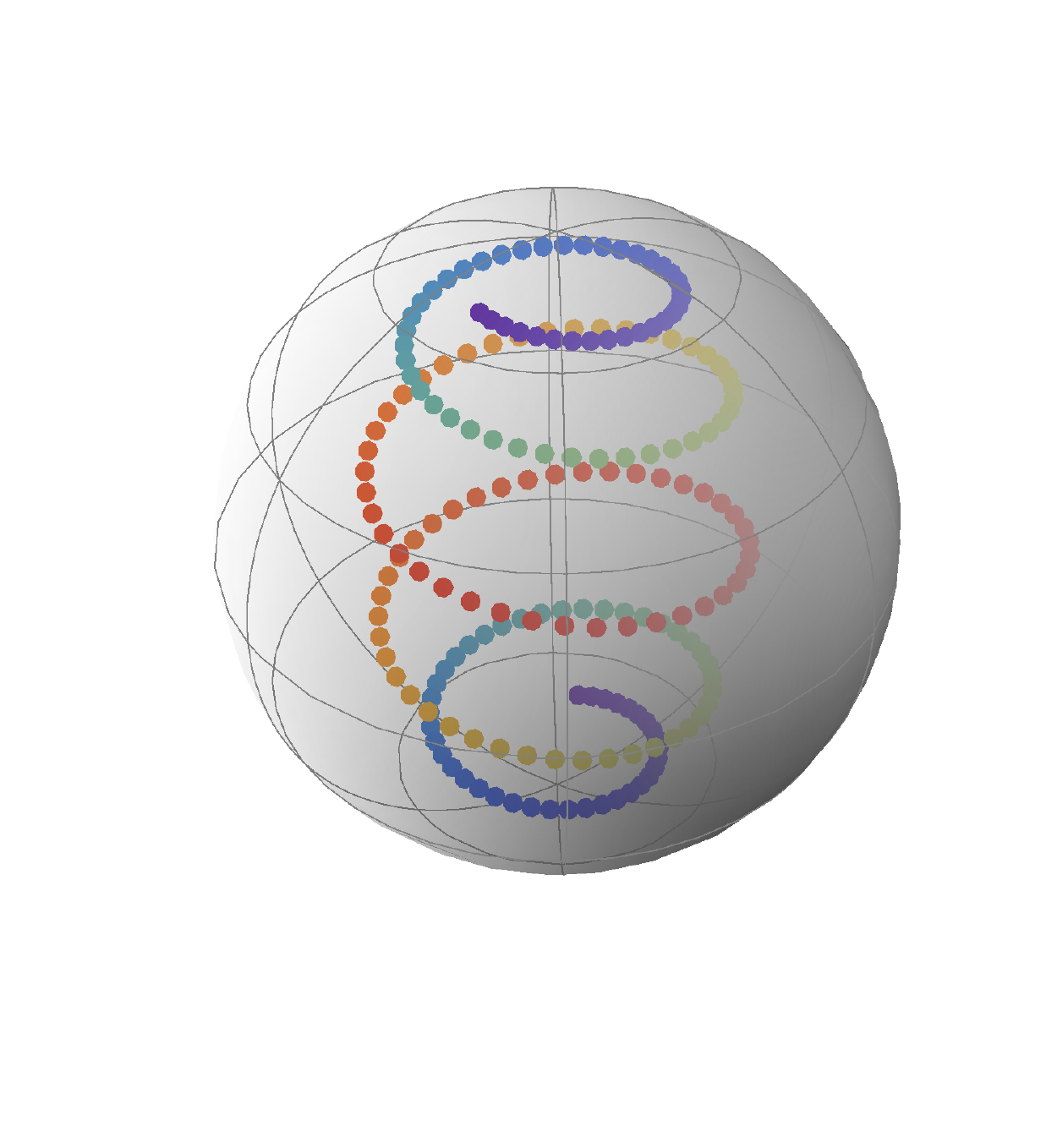}\\{\includegraphics[width=0.5\columnwidth]{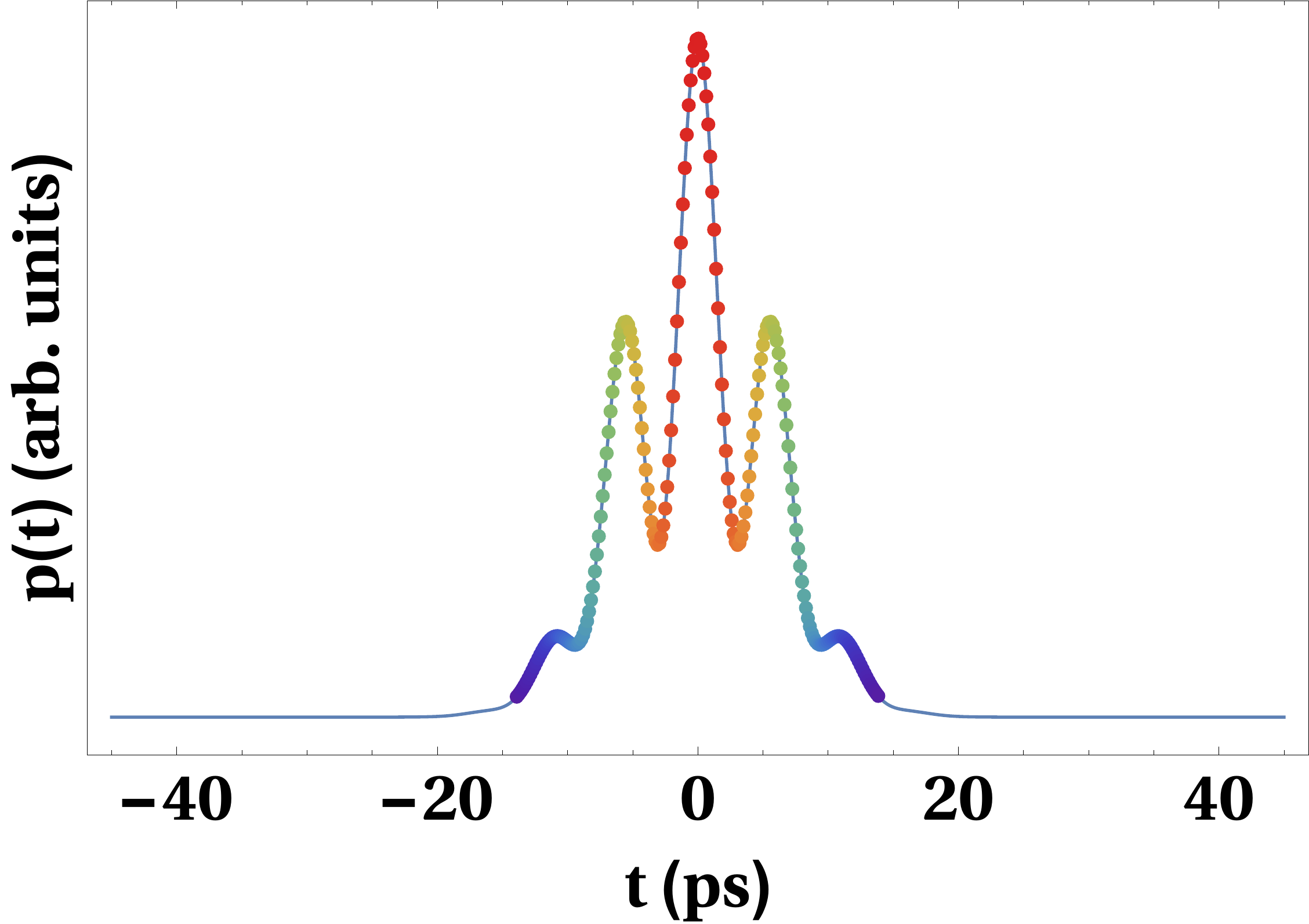}}}&\centered{\includegraphics[width=0.45\columnwidth]{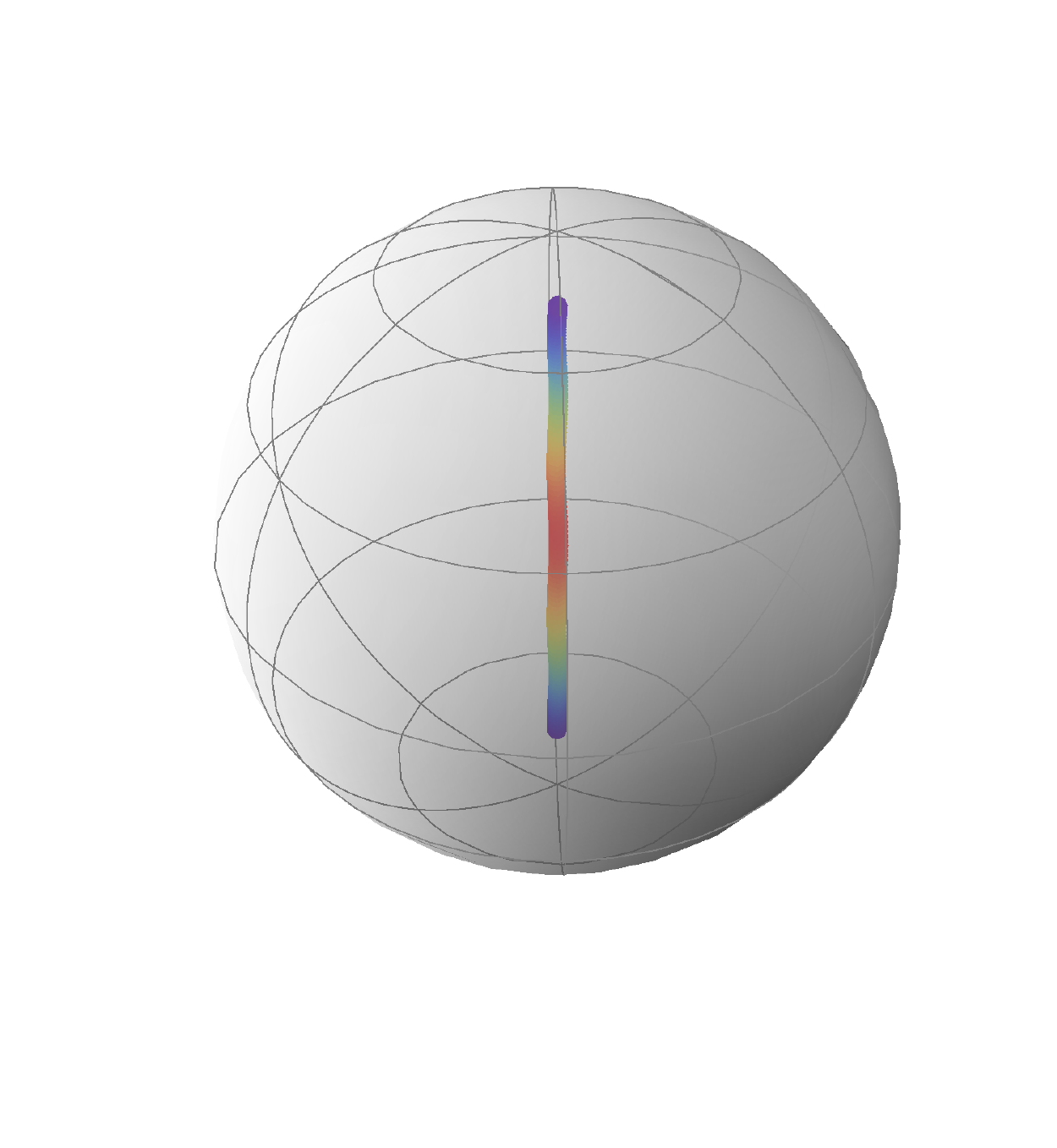}\\{\includegraphics[width=0.5\columnwidth]{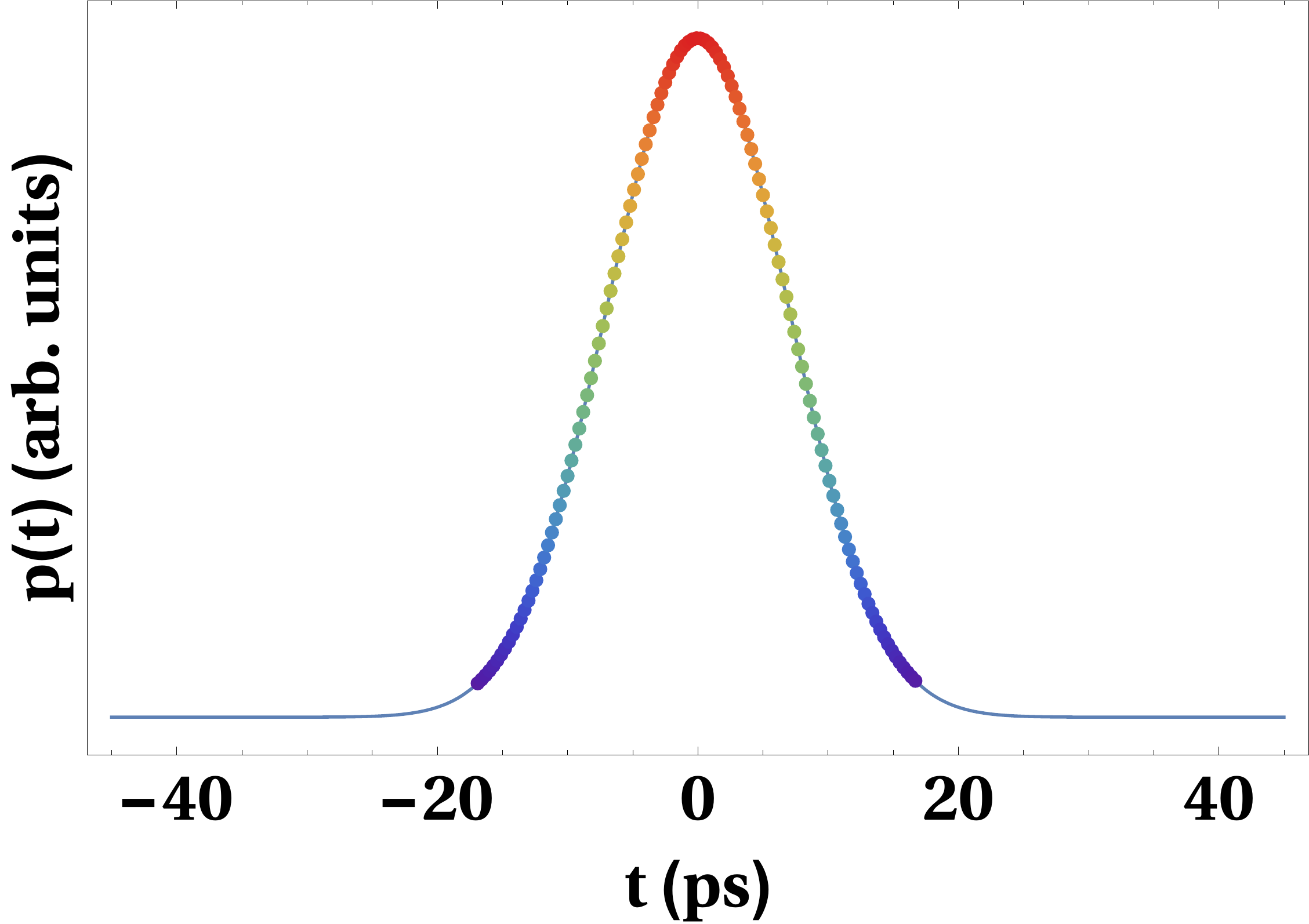}}}\\\hline
			\centered{$500$ m}&\centered{\includegraphics[width=0.45\columnwidth]{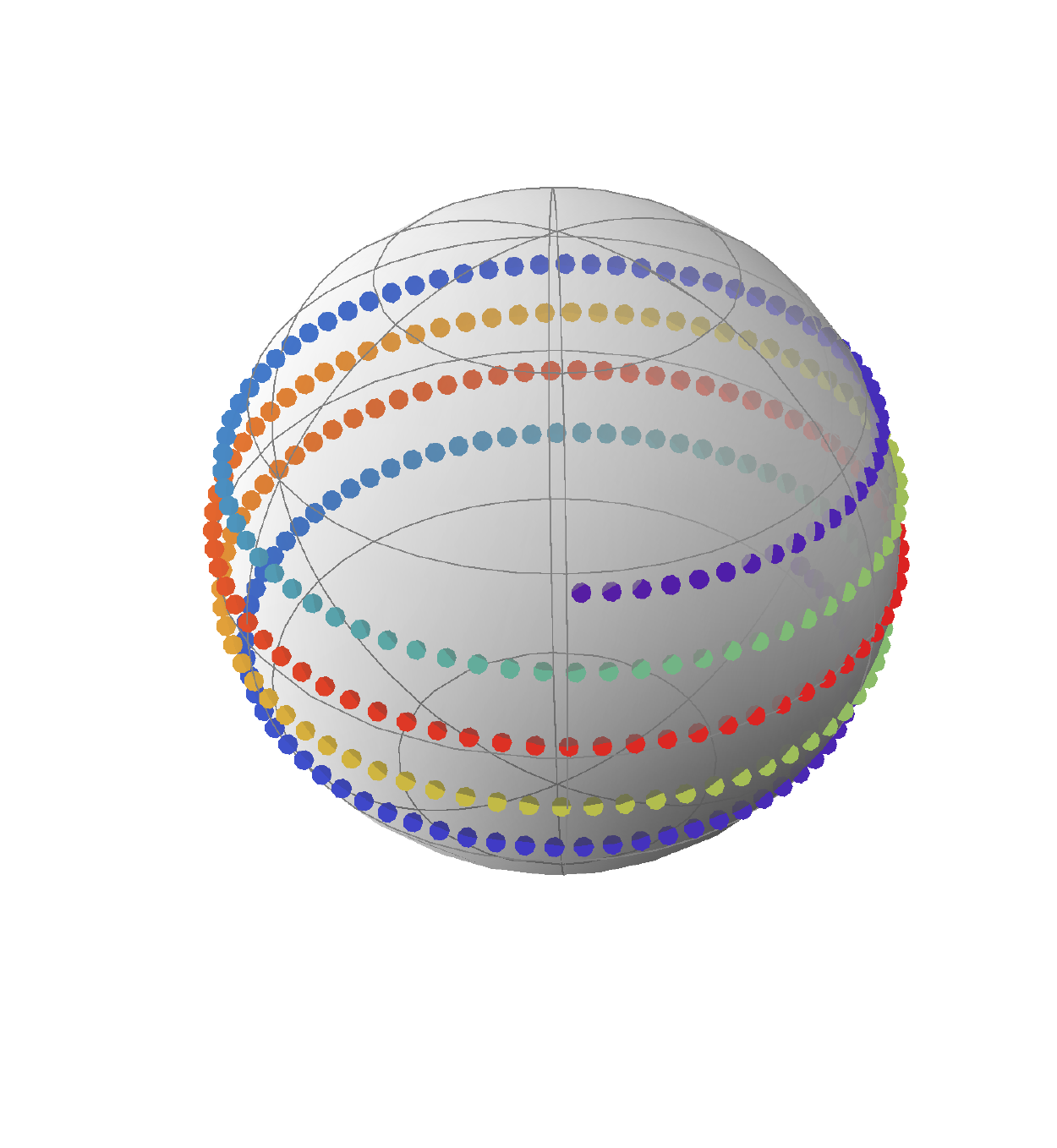}\\\includegraphics[width=0.5\columnwidth]{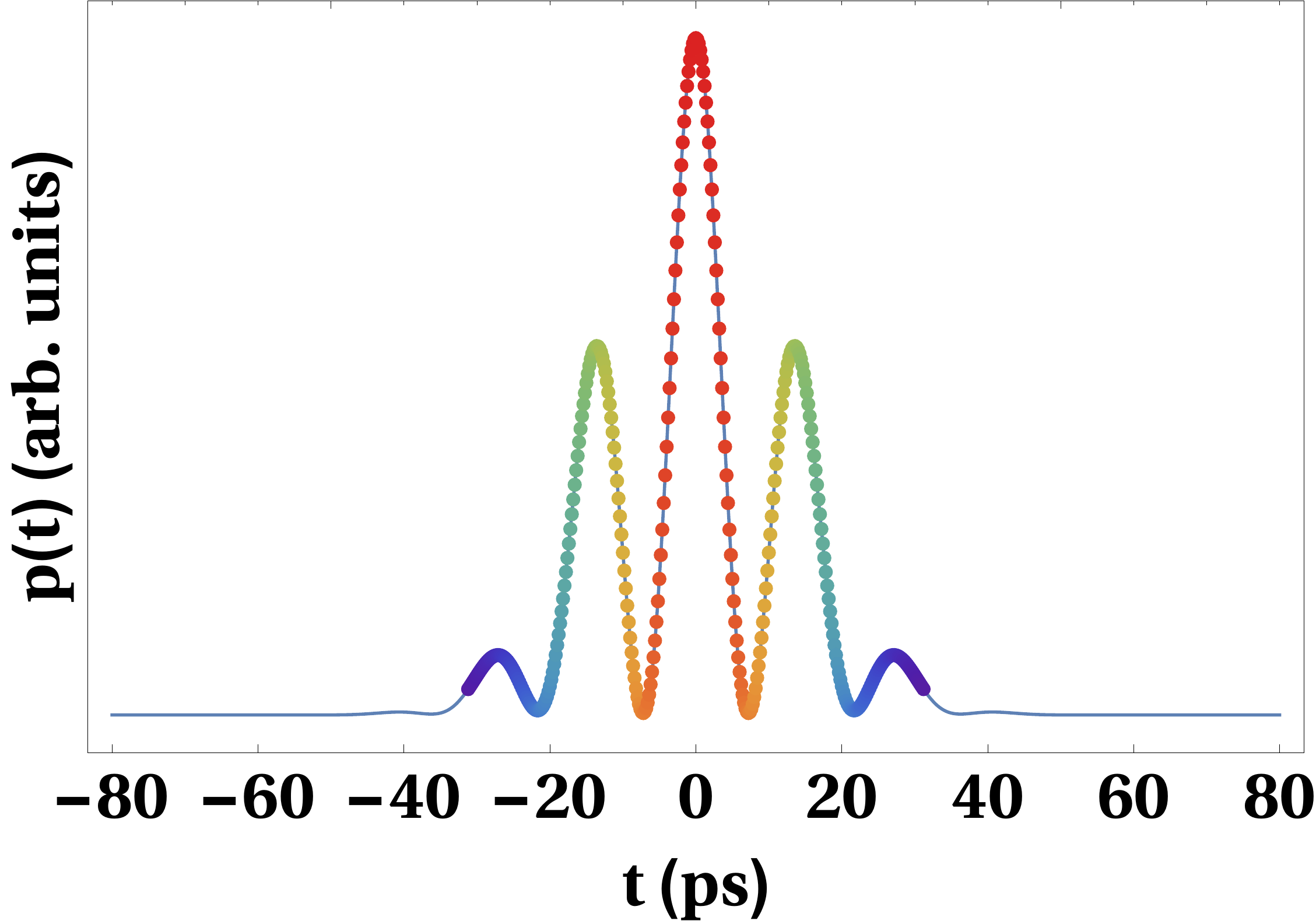}} &
			\centered{\includegraphics[width=0.45\columnwidth]{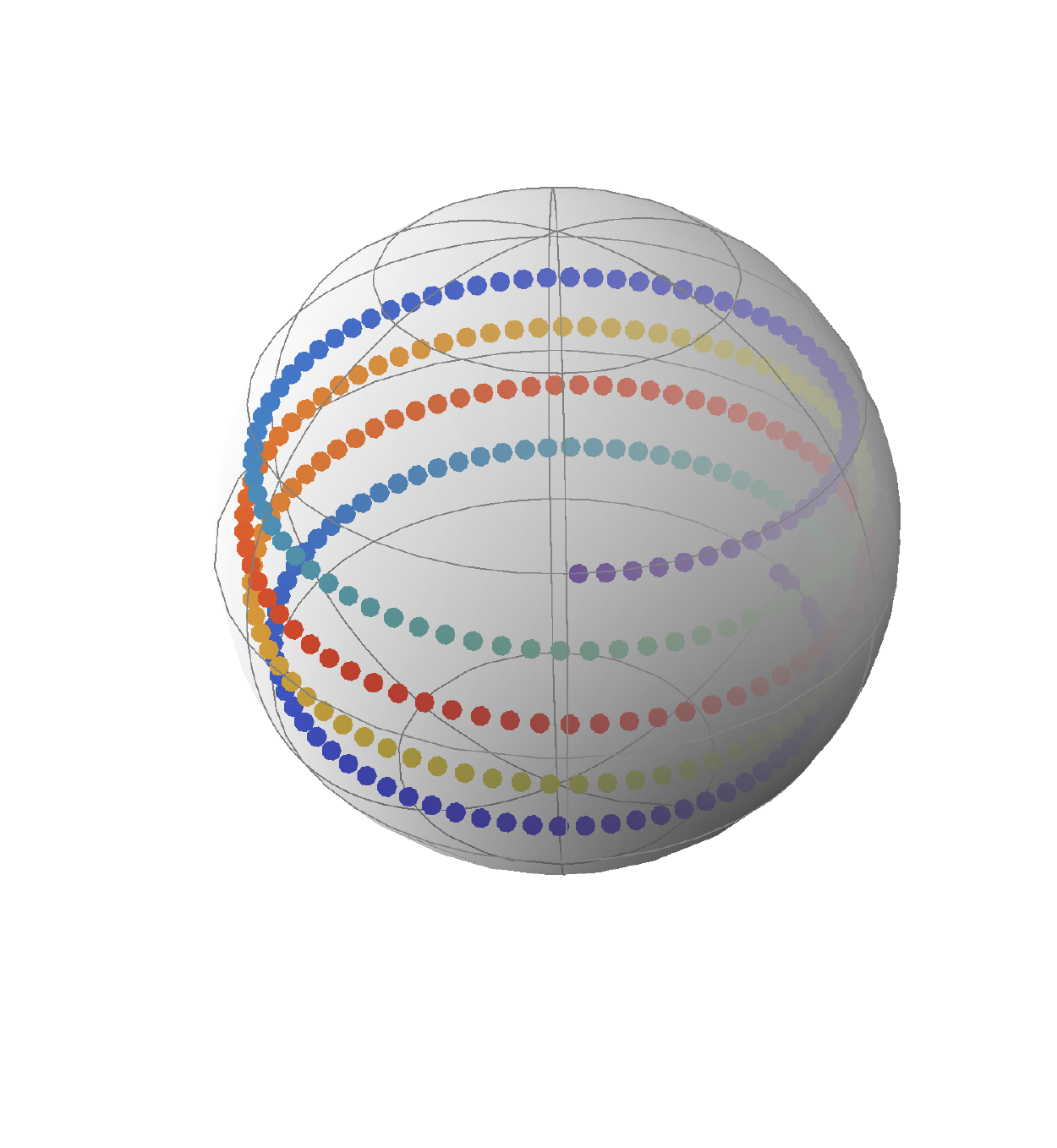}\\\includegraphics[width=0.5\columnwidth]{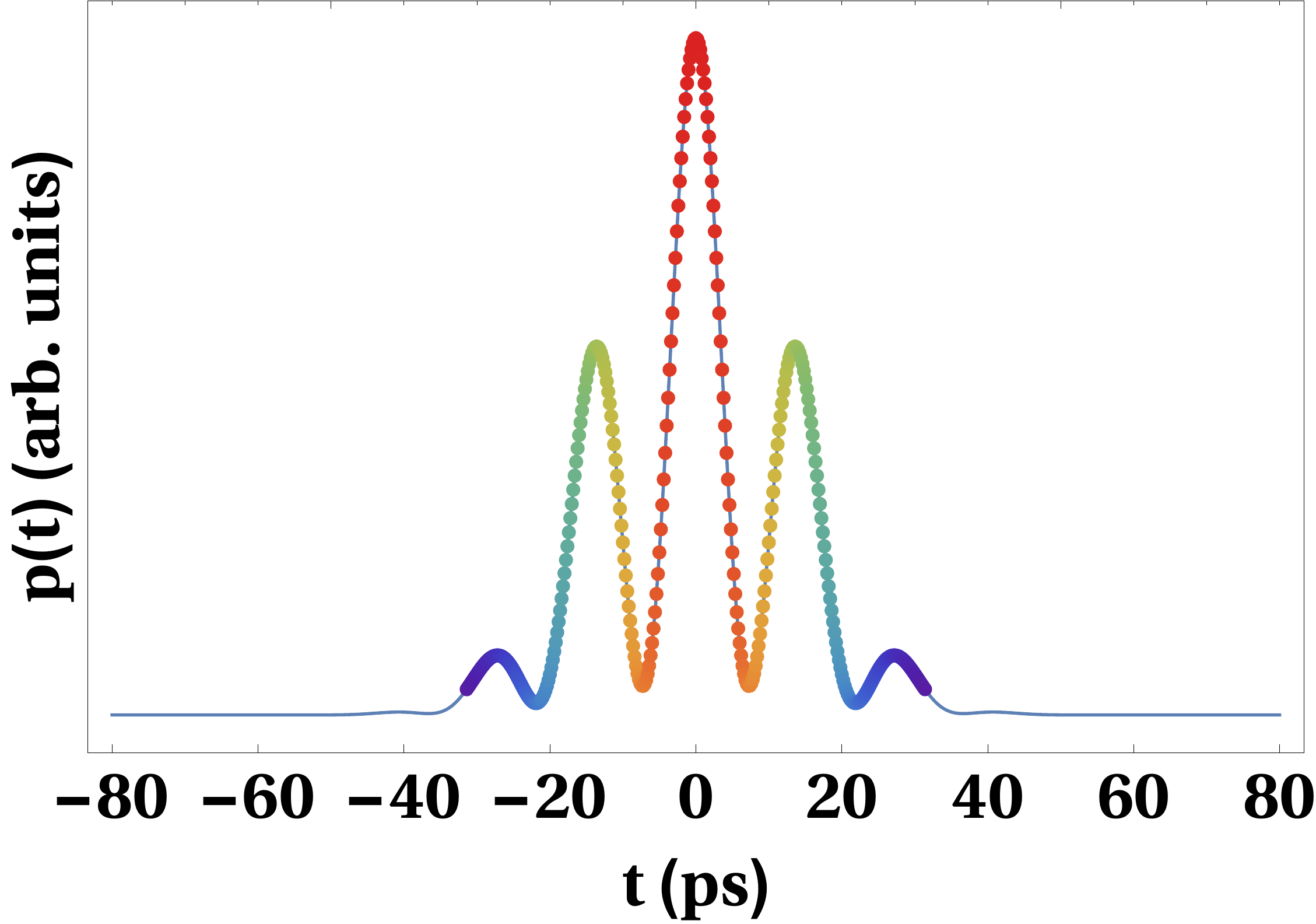}}&\centered{\includegraphics[width=0.45\columnwidth]{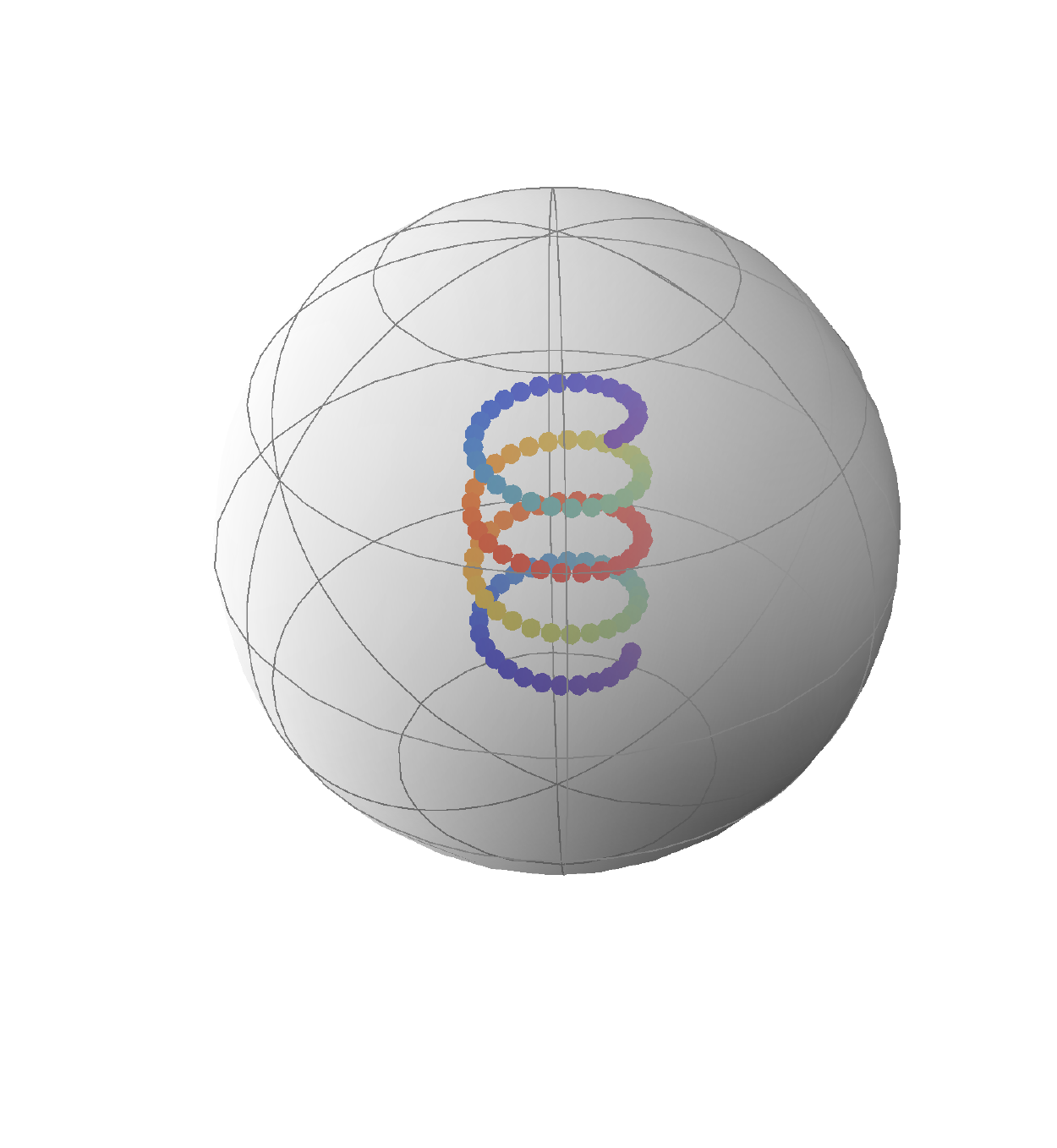}\\\includegraphics[width=0.5\columnwidth]{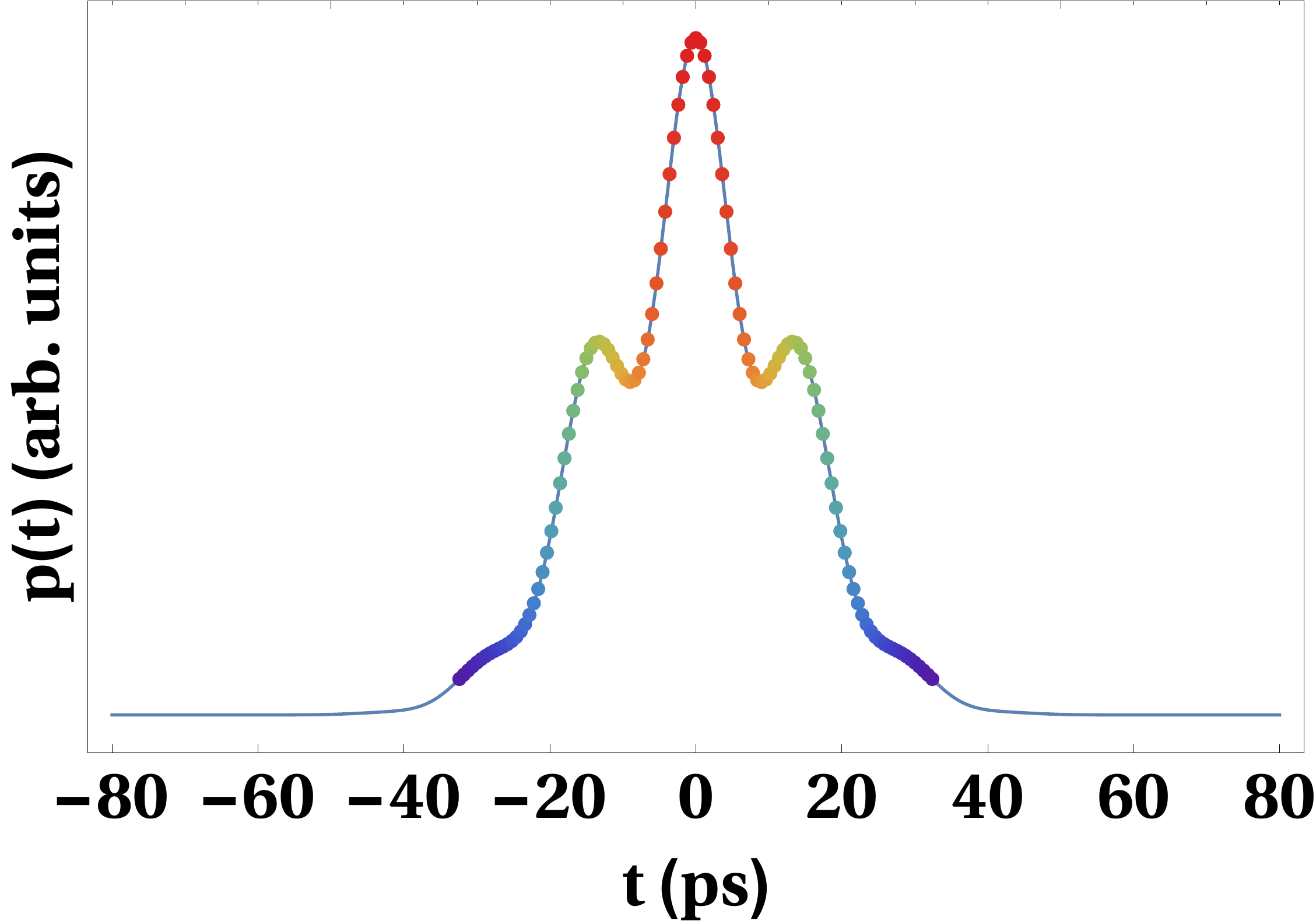}}\\
		\end{tabular}
		\caption{Visualization of a POVM set on the Bloch sphere in the ideal scenario (first column) and in the cases of the detector jitter of $1$ ps (second column) and  $4$ ps (third column). Under each Bloch sphere we put the probability density, $p(t)$, given by \eqref{eq:prob:dens} after propagation through a fiber link by an example input state $\ket{\psi_{in}}=\frac{1}{\sqrt{2}}\left(\ket{0}+\ket{1}\right)$. The probability density for the orthogonal state to $\ket{\psi_{in}}$ shows complementary fringes (maxima are replaced by minima and vice versa). In the case of large detector timing jitter the fringes are not visible. In the rows we put different values of the fiber length. The color in the temperature mapping represents the probability density of the measurement, $\mu(t)$ of the respective POVM element, where the most probable is colored with red (the highest $\mu(t)$).}
		\label{fig:Mg}
	\end{figure*}

	The POVM set can be visualized using the Bloch sphere to represent states $\ket{\psi_M(t)}$ as points and the measurement weights, $\mu(t)$, by assigning color to the respective points using a temperature scaling. An example, for a typical telecom fiber (SMF28e+), can be seen in the first row in \figref{fig:Mg}.  The fiber and wave packet  parameters are the following: $\beta = -1.15 \times 10^{-26}\ \frac{\text{s}^2}{\text{m}}$, $\sigma= 0.65$ ps and $\tau=5$ ps throughout the paper. The points corresponding to the measurement time instants form a spiral on the Bloch sphere. The plot shows  a discrete set of time instants for which the POVM probability, $\mu(t)$, is greater than $5$ \%. The spiral is more squashed for longer fibers as seen by comparing the pictured POVM for fiber lengths $L=200$ m and $L=500$ m. Note that under each Bloch sphere we simulated the outcome of photon arrival time detection for different values of the length of the link $L$ and detector jitter $\sigma_D$ (defined in \eqref{eq:jitter}) for input state $\ket{\psi_{in}}=\frac{1}{\sqrt{2}}\left(\ket{0}+\ket{1}\right)$. This is a simulation of an example measurement result for quantum state reconstruction, which will be analysed later.
	
	In practice, single photon detection systems feature timing jitter, which is an uncertainty in the measured arrival time of the particle.  A timing uncertainty of the detection process can be modelled by convoluting the real probability distribution, $p(t)$ given in \eqref{eq:prob:dens}, with a Gaussian distribution defined by
	\begin{equation}
	q_D(t)=\frac{\exp \left(-\frac{t^2}{2 \sigma_D^2}\right)}{\sqrt{2 \pi  \sigma_D^2}},
	\label{eq:jitter}
	\end{equation}
	where $\sigma_D$ is the timing jitter. This parameter in the case of superconducting nanowire single-photon detectors (SNSPDs) is of the order of $25$ ps \cite{Divochiy2018,Shcheslavskiy2016} and for state-of-the-art ones can reach $1$ ps \cite{Korzh2020}. The POVM, when taking imperfect detectors into account, can then be written as
	\begin{equation}
	\hat{M}_D(t)=\int_{-\infty}^{\infty} \hat{M}(t')\ast q_D(t-t') \dd t'.
	\label{eq:Mt:D}
	\end{equation}
	The measurement operator $\hat M_D(t)$ can be decomposed in terms of mixed states as opposed to $\hat M(t)$, which is defined using pure states, see \eqref{eq:Mt}. This is illustrated in \figref{fig:Mg} where the second and third column show the impact of the detector imperfection. Even a very small jitter increases the entropy of the POVM significantly. Note that for $4$ ps the spiral is degenerated to a line. This effect, however, can be partially compensated for by adding more dispersion (increasing the length of the fiber) as can be seen by comparing the POVMs in the last column. In general, a larger detector timing jitter makes the measurement points collapse on the Bloch sphere, reducing the purity of the POVM elements. A longer fiber, on the other hand, results in a spiral with a greater radius, but the measurement points are localized nearer the equator. This means that, for very long fibers, one would be able to estimate only the phase of the state.

	A dispersive medium, a fiber in the particular realization described above, and time-resolved single photon measurement technique define an informationally complete POVM, as presented in \eqref{eq:Mt} and illustrated in \figref{fig:Mg}. In practical realization, one must decide on a number of time instants to be taken into account,  which will result in the same number of measurements. The difference between this method and a classic projective measurement approach is that here we do not have to change measurement basis to get full information for tomography. Our approach is analogical to a classic setting, where detectors monitor six output ports of measurement apparatus designed such that each port corresponds to one of the four linear and two circular polarizations \cite{Born1980}.
		
	\subsection{Qudit}
\begin{figure}[h!]
		\centering
		\begin{tabular}{c|c}
			\backslashbox[12mm]{\color{black}$L$}{\color{black}$\sigma_D$}&\centered{$0$ ps}\\\hline
			\centered{$200$ m}&\centered{\includegraphics[width=0.45\columnwidth]{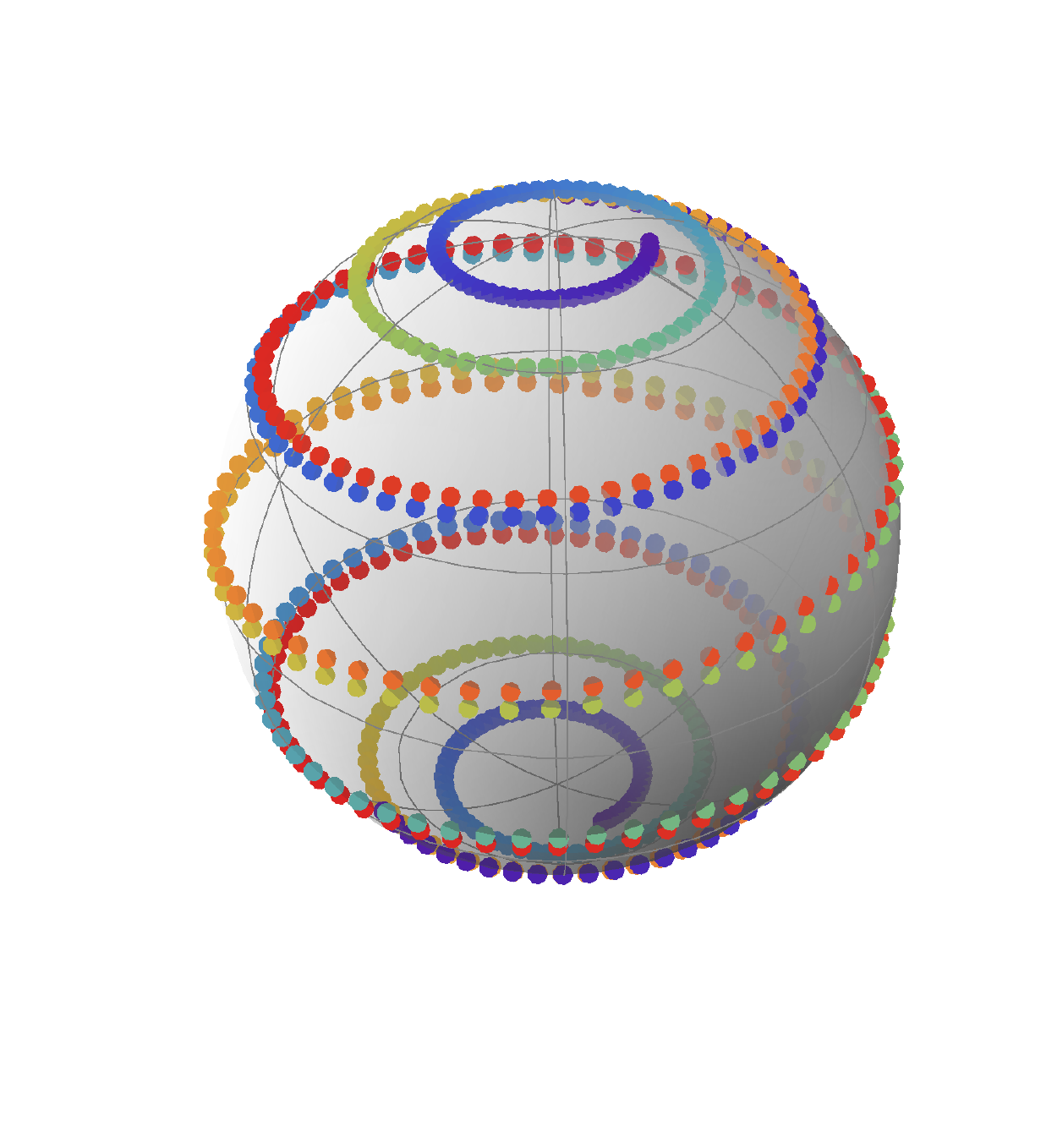}} \\ \hline
			\centered{$500$ m}&\centered{\includegraphics[width=0.45\columnwidth]{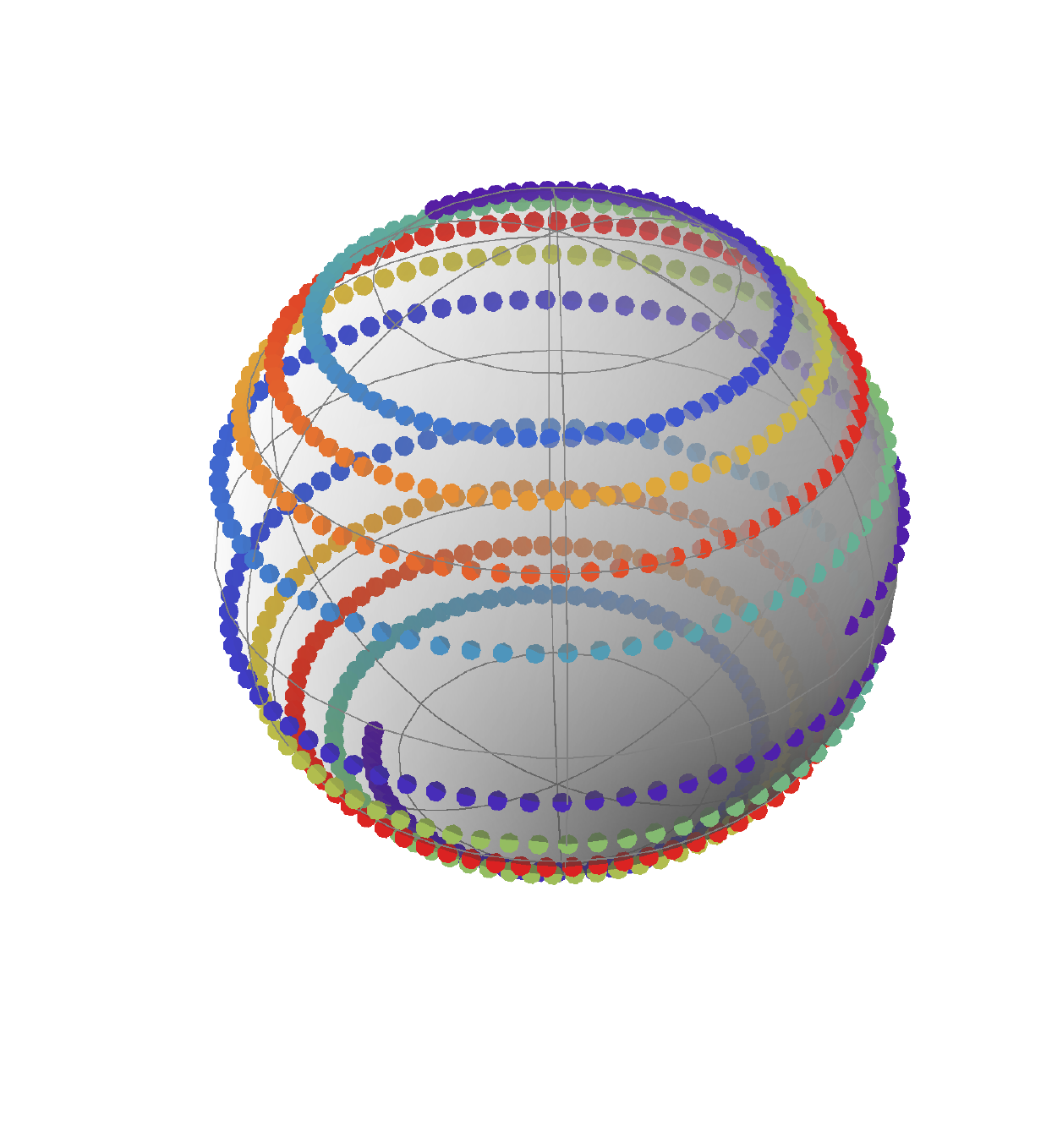}}\\
		\end{tabular}
		\caption{Illustration of pairs of measurements points on the Bloch sphere for qutrits in the ideal scenario. In the rows we put different values of length of the fiber. The highest probabilities of our detector registering a photon are marked as red and the lowest probabilities in violet.}
		\label{fig:Mgq3}
	\end{figure}

	The framework introduced for qubits can be easily generalized to qudits. We define a qudit as a photon delocalized in $d$ wave packets separated by a time interval $\tau$. A wave function for a qudit can be defined, in analogy to \eqref{eq:def:qubit:cont}, as
	\begin{equation}
	\psi(t)=\sum_{n=0}^{d-1}\alpha_n u\left(t-n\tau +\frac{d-1}{2}\tau\right),
	\end{equation}
	where $\alpha_n$ are complex numbers satisfying the normalization condition, $\sum_{n=0}^{d}|\alpha_n|^2=1$, and $u(t)$ is given by the \eqref{eq:ut}.
	The basis vectors are defined by the following formula 
	\begin{equation}
	\ket{n}=\int_{-\infty}^{\infty}\dd t\, u\left(t-n\tau+\frac{d-1}{2}\tau\right)\ket{t},
	\label{eq:def:qudit}
	\end{equation}
	where $n=0,\dots, d-1$. The overlap of two arbitrary states reads
	\begin{equation}
	\braket{n|k}=e^{-\frac{\tau ^2 (k-n)^2}{16 \sigma ^2}}.
	\end{equation}
	In turn, the measurement operator as defined by \eqref{eq:Mt} can be generalized by using the weight
	\begin{equation}
	\mu(t)=\frac{\sigma}{{\sqrt{\pi } \sqrt{C}}}\left(e^{-\frac{\sigma ^2 (t+\frac{d-1}{2}\tau )^2}{C}}+\ldots+e^{-\frac{\sigma ^2 (t-\frac{d-1}{2}\tau )^2}{C}}\right),
	\end{equation}
	and measurement vector
	\begin{equation}
	\ket{\psi_M(t)}=\frac{1}{\sqrt{\mu(t)}}\left(
	\begin{array}{c}
	u_L(t+\frac{d-1}{2}\tau) \\
	u_L(t+\frac{d-3}{2}\tau) \\
	\vdots\\
	u_L(t-\frac{d-3}{2}\tau) \\
	u_L(t-\frac{d-1}{2}\tau)\\
	\end{array}
	\right).
	\label{eq:qudit:povm:gen}
	\end{equation}
	It can be shown that the completeness of the measurement operators also holds, $\int_{-\infty}^{\infty} \hat{M}(t)\dd t \approx \mathbb{1}$, with the assumption of the approximate orthogonality of the basis states. Next, the detector jitter can be taken into account in the same way as before, cf. \eqref{eq:Mt:D}. Similarly as for qubits, but only for the case with no detector jitter, the POVM can be represented on the Bloch ball by using the Majorana representation \cite{Majorana1932, Kolenderski2010}. An example for a qutrit is shown in \figref{fig:Mgq3} and commented in the next section.
	
	\section{Quantum state tomography}
	
	We postulate that a large number of identical copies of a given state is generated. The state is then reconstructed based on the statistics of the temporal detections, see example of $p(t)$ in \figref{fig:Mg}. In an experiment, it is challenging to obtain a perfect informationally-complete set of measurement operators. Therefore, we need to evaluate the efficiency of realistic measurement operators.

	We assume that we are able to describe the imperfections of the experimental apparatus and therefore, we apply the formula for the detection probability which contains the detector jitter. The measurement operators, which were discussed in the previous section, constitute an approximate POVM and can be used as a source of information for quantum state tomography. Mathematically, we follow the Born rule to describe the detection probabilities. Expected photon counts are then computed, assuming that we use $10^3$ identical photons. Since our POVM as given by \eqref{eq:Mt:D} is defined in the time domain, we select a discrete subset of $26$ measurement operators ($25$ in the case of entangled photons).

	In reality, measurement results are burdened with errors and noise. We consider the Poisson noise, which is a typical form of uncertainty associated with the measurement of photons \cite{Hasinoff2014}. Since our model is based on photon counting, the Poisson noise appears to be adequate. Thus, we numerically generate a set of experimental data by imposing the Poisson noise on the expected photon counts. For any input state this approach allows us to simulate realistic experimental situation.
	
	Then, we employ two very widely used quantum state tomography techniques: the maximum likelihood estimation (MLE) \cite{Hradil1997,Banaszek1999} and the least squares (LS) method \cite{Opatrny1997}, which are often compared in terms of their efficiency, e.g. \cite{Acharya2019}. For the convenience of numerical analyses we adopt from Ref.~\cite{James2001} the factorization of the unknown density matrix, which provides Hermiticity, positivity and normalization:
	\begin{equation}\label{eq:kwiat}
	\rho =\frac{ W^{\dag} W }{\tr \{W^{\dag} W\}},
	\end{equation}
	where $W$ denotes a complex left triangular matrix. This decomposition ensures that the estimated matrix is physical and belongs to the quantum state set. The quality of the reconstructed state is quantified by computing the quantum fidelity \cite{Jozsa1994}, which is defined for two mixed states $\rho$, $\sigma$ by
	\begin{equation}\label{statesfidelity}
	\mathcal{F}(\rho, \sigma) := \left( \tr \sqrt{\sqrt{\rho} \,\sigma\,\sqrt{\rho}} \right)^2.
	\end{equation}
	It is easy to verify that $0 \leq \mathcal{F}(\rho, \sigma) \leq 1$ and $\mathcal{F}(\rho, \sigma)=1$ if and only if $\rho = \sigma$. Furthermore, it is symmetric, i.e. $\mathcal{F}(\rho, \sigma) = \mathcal{F}(\sigma, \rho)$, which does not stem straightforwardly from the definition. That formula can be simplified if one considers only pure states, i.e. $\mathcal{F}(\rho, \sigma) = | \braket{\psi_{\rho}|\psi_{\sigma}}|^2$ for $\rho=\ket{\psi_{\rho}}\bra{\psi_{\rho}}$ and $\sigma =\ket{\psi_{\sigma}}\bra{\psi_{\sigma}}$.
	
	We use quantum state fidelity $\mathcal{F}(\rho_{in},\rho_{out})$ to evaluate the quality of our tomography framework. Since the value of the fidelity depends on the initial density matrix $\rho_{in}$, we introduce the \textit{average fidelity}, $\mathcal{F}_{av}$ as the figure of merit. It is defined as the mean value computed over all possible input states, $\rho_{in}$. It allows us to determine the average performance of our quantum tomography scheme. In practice, we cannot find analytically the average fidelity over the entire state set, so we select a representative sample of quantum states for numerical analysis. The sample is selected based on a parametric-dependent structure of the density matrix where each parameter goes over the full range. Then, each input state from this discrete set is sent through the fiber, next we simulate measurement results distorted by the Poisson noise and finally one can reconstruct the density matrix.
	
	\subsection{Qubit}

	For a qubit, the triangular matrix, $W$, parametrizing the density operator, $\rho$, given by \eqref{eq:kwiat} takes the following form:
	\begin{equation}
	W =\begin{pmatrix} w_1 & 0 \\ w_3 + i \,w_4 & w_2 \end{pmatrix},
	\end{equation}
	where $w_1,w_2,w_3,w_4 \in \mathbb{R}$. Thus, the problem of reconstructing the initial density matrix can be formulated in terms of determining the values of $w_1, w_2, w_3, w_4$.
	
	\begin{table}[h]
		\begin{tabular}{c|c|c|c|c}
			\multicolumn{2}{c|}{{
					\backslashbox[24 mm]{\color{black}$L$}{\color{black}$\sigma_D$}}} & $0$ ps & $1$ ps & $4$ ps \\
			\hline
			\multirow{2}{*}{200 m} & LS &$0.9995(12)$  & $0.9982(18)$ & $0.61(18)$ \\ \cline{2-5} 
			& MLE & $0.998(2)$ & $0.987(11)$  & $0.58(18)$  \\ \hline
			\multirow{2}{*}{500 m} & LS & $0.996(4)$ & $0.9962(34)$ & $0.9951(45)$ \\ \cline{2-5}
			& MLE & $0.9921(31)$ & $0.9897(42)$ & $0.91(4)$
		\end{tabular}
		\caption{Average fidelity with standard deviation for quantum tomography of qubits computed numerically for different values of experimental parameters. The results were obtained over a sample of $9261$ qubits. The experimental results were simulated with the Poisson noise. The same set of data was used for both LS and MLE quantum tomography techniques.}
		\label{fig:qubit}
	\end{table}
	
	Numerical simulation has been conducted to test the effectiveness of our measurement operators for quantum tomography in different experimental setups. The main results are gathered in \tabref{fig:qubit}. One can observe that for $\sigma_D = 0$ (measurements without jitter) both quantum tomography methods result in an average fidelity very close to $1$. For both values of the fiber length the figures are very close (we consider the difference negligible). This outcome confirms that in the ideal scenario (no jitter) any quantum state can be reconstructed flawlessly and the Poisson noise does not reduce the average fidelity. Both MLE and LS methods lead, on average, to the relevant quantum state.
	
	If we analyze the results in the rows of \tabref{fig:qubit}, one can easily notice that for the fiber length of $200$ m the average fidelity decreases as the detector jitter increases. It is a consequence of the detector jitter leading to a greater uncertainty. However, when the detector jitter equals $1$ ps both quantum tomography methods can still reconstruct the initial state with high accuracy.
	
	The most interesting conclusion can be drawn if we compare the results in the columns. One can see that when the detector jitter is fixed at $4$ ps, we obtain significantly higher average fidelity for the longer fiber. In the case of $L=200$ m the state reconstruction appears highly inaccurate since we have low average fidelity. However, it can be improved if we extend the length of the fiber to $500$ m. This means that the two parameters, the length of a fiber and the detector jitter, have opposing impacts on the average fidelity. One can reduce the errors due to the detector jitter by using a longer fiber. Numerical results are in agreement with \figref{fig:Mg}, where the measurement points are depicted on the Bloch ball. For the fixed jitter the operators $\hat{M}_D (t)$, which constitute the POVM, are less mixed if the fiber is longer.
	
	\subsection{Qutrit}
	
	The basis states for a qutrit are defined as in \eqref{eq:def:qudit} and the states generating the POVM are given by \eqref{eq:qudit:povm:gen} for $n=3$. To visualize the POVM on the Bloch sphere for qutrits  we utilize Majorana representation \cite{Majorana1932}. It allows to associate a pair of two-dimensional states with a qutrit. For the qutrit measurement vector, given by \eqref{eq:qudit:povm:gen}, one can write a quadratic Majorana polynomial $p(\ket{\psi_M})=0$ \cite{Kolenderski2010}:
	\begin{equation}
	e^{\frac{i (t-\tau )^2}{4 \beta  L-2 i \sigma ^2}}z^2-\sqrt{2} e^{\frac{i t^2}{4 \beta  L-2 i \sigma ^2}}z +
	e^{\frac{i (t+\tau )^2}{4 \beta  L-2 i \sigma ^2}}=0.
	\label{eq:Mpol}
	\end{equation}
	Stereographic projection is used to associate the two complex numbers with two points on the Bloch sphere. We visualize this for two different lengths of the fiber to see how the pairs of points are distributed on the Bloch sphere. 
	
	In the case of qutrits we follow a very similar quantum tomography procedure as for qubits. First, we use the same parametric-dependent formula for the unknown density matrix given by \eqref{eq:kwiat}. Here, the matrix $W$ depends on $9$ real parameters:
	\begin{equation}
	W =\begin{pmatrix} w_1 & 0 & 0 \\ w_4 + i \,w_5 &  w_2 & 0 \\  w_8 + i\, w_9 & w_6 + i\, w_7 & w_3 \end{pmatrix}.
	\end{equation}
	Thus, the problem of state reconstruction for qutrits can be translated into finding the values of $w_1, w_2, \dots, w_9$. Next, to evaluate the effectiveness of the POVM for qutrits, we follow exactly the same steps as for qubit. 
	
	\begin{table}[ht]
		\begin{tabular}{c|c|c|c|c}
			\multicolumn{2}{c|}{{
					\backslashbox[24 mm]{\color{black}$L$}{\color{black}$\sigma_D$}}} & $0$ ps & $1$ ps & $4$ ps \\
			\hline
			\multirow{2}{*}{$200$ m} & LS & $0.988(13)$ & $0.949(52)$ & $0.61(24)$ \\ \cline{2-5} 
			& MLE & $0.988(15)$ & $0.959(36)$ & $0.61(24)$ \\ \hline
			\multirow{2}{*}{$500$ m} & LS & $0.971(31)$ & $0.969(31)$ & $0.91(8)$ \\ \cline{2-5} 
			& MLE & $0.970(29)$ & $0.969(29)$ & $0.91(8)$         
		\end{tabular}
		\caption{Average fidelity with standard deviation for quantum tomography of qutrits computed numerically for different values of experimental parameters. The results were obtained over a sample of $9261$ qutrits. The experimental results were simulated with the Poisson noise. The same set of data was used for both LS and MLE quantum tomography techniques.}
		\label{tab:qutrit}
	\end{table}

	The results of the average fidelity for quantum tomography of qutrits are gathered in \tabref{tab:qutrit}. One can observe that in the case of perfect measurements (no jitter), both methods can reconstruct the initial quantum state with only limited accuracy. If we analyze the results in the row for $L=200$ m, we can observe a substantial influence of the detector jitter on the average fidelity. Particularly, if $\sigma_D = 4$ ps the average fidelities are relatively small, but they are very close to the results in the case of qubit reconstruction. The difference is that for qutrits we get a higher standard deviation, which means that the variation of the set of quantum state fidelities is greater than for qubits.

	Two conclusions can be drawn from the results in the columns. First, if $\sigma_D = 1$ ps or $\sigma_D = 4$ ps, the MLE and LS methods appear to follow the same tendency as for qubits, i.e. the average fidelity increases when we use a longer fiber. Second, it should be noted that for $\sigma_D = 4$ ps the scale of improvement is most significant. Interestingly, both quantum state tomography techniques yielded the same results. However, the LS method leads to worse precision than in the case of qubits.

\subsection{Entangled qubits}

To demonstrate the performance of our quantum tomography framework on entangled photon pairs, we shall consider input states in the form
\begin{equation}\label{input}
\ket{\Phi^+} = \frac{1}{\sqrt{2}} \left( \ket{00} + e^{i \phi} \ket{11}  \right),
\end{equation}
where $\phi$ is the relative phase ($0\leq\phi< 2 \pi$). This type of two-photon entangled state is commonly considered in quantum communication protocols based on time-bins since it can be produced by spontaneous four-wave mixing (SFWM) in a dispersion shifted fiber \cite{Takesue2005,Takesue2009}, by spontaneous parametric down conversion (SPDC) \cite{Marcikic2004} as well as by a source utilizing quantum dots \cite{Jayakumar2014,Versteegh2015}. 

The problem of relative phase estimation was first solved for polarization entangled photons \cite{White1999}. In the case of time-bin entangled qudits it has recently been undertaken \cite{Nowierski2016,Ikuta2017} though by different measurement techniques. For such states, as given by \eqref{input}, we simulated measurement results which are distorted by the Poisson noise. We use $25$ measurement operators defined as
\begin{equation}\label{enoperator}
M (t_i, t_j) := \hat{M}_D (t_i) \otimes \hat{M}_D (t_j),
\end{equation}
where $\hat{M}_D (t_i)$ denotes the single qubit measurement operator with the detector jitter \eqref{eq:Mt:D} and $t_i, t_j$ belong to a discrete $5-$elements subset selected from time domain.

Numerical data is used to perform quantum state tomography by MLE and LS method. We assume that there is no \textit{a priori} knowledge about the system, i.e. the unknown quantum state takes the general form of $4\times 4$ density matrix. Thus, we follow \eqref{eq:kwiat}, where the matrix $W$ depends on $16$ real parameters
\begin{equation}
W = \begin{pmatrix} w_1 & 0 & 0 &0 \\ w_5 + i\, w_6 &  w_2 & 0 &0 \\  w_{11} + i \,w_{12} & w_7 + i\, w_8 & w_3 &0 \\ w_{15} + i\, w_{16} & w_{13} + i\, w_{14} & w_9 + i\, w_{10} & w_4 \end{pmatrix}.
\end{equation}
We consider a sample of $200$ input states defined as \eqref{input} with different values of the relative phase. For each input state we generate a set of realistic measurement results and then we introduce this data to MLE and LS algorithms in order to estimate the values of $16$ parameters. Finally, we compute the fidelity between the reconstructed state and the original state $\rho_{in} = \ket{\Phi^+} \bra{\Phi^+}$. 

 \tabref{tab:relativephase} presents the values of the average state fidelity computed for different combinations of experimental parameters.
\begin{table}[h]
\centering
	\begin{tabular}{c|c|c|c|c}
		\multicolumn{2}{c|}{{
				\backslashbox[24 mm]{\color{black}$L$}{\color{black}$\sigma_D$}}} & $0$ ps & $1$ ps & $4$ ps \\
		\hline
		\multirow{2}{*}{$200$ m} & LS & $0.974(15)$ & $0.937(41)$ & $0.48(31)$ \\ \cline{2-5} 
		& MLE & $0.988(1)$ & $0.951(36)$ & $0.45(34)$ \\ \hline
		\multirow{2}{*}{$500$ m} & LS & $0.981(1)$ & $0.978(11)$ & $0.85(1)$ \\ \cline{2-5} 
		& MLE & $0.99(1)$ & $0.984(16)$ & $0.85(1)$         
	\end{tabular}
	\caption{Average fidelity with standard deviation for entangled qubits tomography, computed numerically for different values of experimental parameters. Each value was obtained over a sample of $200$ states. The experimental results were simulated with the Poisson noise. The same set of data was used for both LS and MLE quantum tomography techniques.}
\label{tab:relativephase}
\end{table}
One can observe very similar tendencies as in the earlier examples. For $\sigma_D = 4$ ps and $L= 200$ m the quantum tomography framework appears inefficient, whereas if we replace the length of the fiber with $L = 500$ m the average fidelity improves.

It is worth noting that the length of the fiber can compensate for the detector even for greater values of the timing jitter. If we consider $\sigma_D = 20$ ps and $L= 5$ km, we get the average fidelity equal to $0.85(14)$ (LS) and $0.80(20)$ (MLE). Additionally, these figures can rise if one performs more measurements. By taking $36$ operators of the form \eqref{enoperator}, we can increase the average fidelity to $0.97(2)$ (LS) and $0.88(12)$ (MLE). It appears that the interdependence between the detector jitter and the fiber length should be studied in detail. However, changing the length of the fiber involves adapting the choice of time instants due to dispersion which causes broadening the pulse widths of the photons. Extensive analysis of the average fidelity for the full range of lengths would require more processing power and for this reason it will be the subject of future research.
	
\section{Summary and outlook}
	
	In conclusion, we have demonstrated an effective method of measuring quantum states encoded in temporal modes of photons. Numerical tools allowed us to verify under which circumstances the reconstruction methods are the most effective for a set of realistic experimental parameters. To investigate the performance of our quantum state tomography schemes, we employed the average fidelity. Our analysis indicates that a longer fiber can compensate for the effects caused by detector jitter. 
	
	There are remaining research problems that need further investigation. For instance, it will be useful to perform more numerical simulations of the average fidelity for a wider range of parameters in order to obtain a broad view on the effectiveness of our schemes. Moreover, the quantum tomography framework shall be tested on different multi-photon states, especially more types of entangled photon pairs will be considered.
	
\vspace{1cm}
	\section*{Acknowledgments}

	K.~S.-K. was supported by the National Science Centre, Poland (NCN) (grant no.~2016/23/N/ST2/02133). P.~K. and A.~C. acknowledge the support from the Foundation for Polish Science (FNP) (project First Team co-financed by the European Union under the European Regional Development Fund). 
	
	All the authors acknowledge support by the National Laboratory of Atomic, Molecular and Optical Physics, Torun, Poland.

\end{document}